\documentstyle[aps,prd]{revtex}
\begin{document}

\def\be{\begin{equation}}
\def\ee{\end{equation}}\def\beq{\begin{eqnarray}}
\def\eeq{\end{eqnarray}}\def\s{\sigma}\def\p{\rho}
\def\G{\Gamma}\def\F{_2F_1}\def\an{analytic}
\def\ac{\an{} continuation}\def\hsr{hypergeometric series representations}
\def\hf{hypergeometric function}\def\ndim{NDIM}\def\quarto{\frac{1}{4}}
\def\half{\frac{1}{2}}

\draft
\title{NDIM achievements: Massive, Arbitrary tensor rank and N-loop
insertions in Feynman integrals}   \author{A. T. Suzuki and A. G. M. Schmidt}
\address{Universidade Estadual Paulista -- Instituto de F\'{\i}sica
Te\'orica, R.Pamplona, 145, S\~ao Paulo SP, CEP 01405-900, Brazil. }
\date{\today}
\maketitle
\begin{abstract}
One of the main difficulties in studying Quantum Field Theory, in the
perturbative regime, is the calculation of D-dimensional Feynman 
integrals. In general, one introduces the so-called Feynman parameters
and associated with them the cumbersome parametric integrals. Solving
these integrals beyond the one-loop level can be a difficult task. 
Negative dimensional integration method (\ndim{}) is a technique whereby such 
problem is dramatically reduced. In this work we present the calculation of 
two-loop integrals in three diferent cases: scalar ones with three diferent 
masses, massless with arbitrary tensor rank, with N-insertions of a
2-loop diagram.

\end{abstract}

\section{Introduction.}
Perturbative calculations in Quantum Field Theory are often a hard task,
specially if one does not have a suitable approach to tackle the 
problem. Among the several techniques available in the market the most
popular is  the Feynman parametrization\cite{zuber}; in fact, if one is clever
enough hefty calculations (at 2-loop level) can be done\cite{kramer}.
However, in our point of view, this is not the most adequate nor elegant
method to solve Feynman integrals whether one is considering covariant or
non-covariant gauges.

On the other hand, Chetyrkin {\it et al} developed the integration by 
parts in configuration space (associated with Gegenbauer polynomials) and 
performed (at 4-loops) even heftier calculations\cite{russo}. However, their 
technique has a drawback: if the diagrams have more than two external legs 
manipulation of Gegenbauer polynomials becomes very difficult \cite{kramer} to 
handle.

Mellin-Barnes' contour integration is a third option in the market. Each
propagator is Mellin-transformed\cite{boos} -- a very simple step -- 
and using Barnes' lemmas and summing over the residues, it is possible to 
write the Feynman integrals as hypergeometric functions or  
hypergeometric-like
series. Smirnov\cite{2box} solved the massless double-box with 
Mellin-Barnes approach. Such Mellin integrals are parametric-like integrals
-- though  they are much simpler to solve than Feynman-like ones because Cauchy
theorem can be applied straightforwardly.

\ndim{}\cite{halliday} is a technique where the parametric integrals do 
not appear. We start up with a Gaussian integral, which is well-behaved, 
perform Taylor expansion and solve systems of algebraic equations. All the
calculations can be done analytically and the results are given for 
arbitrary exponents of propagators and space-time dimension $D$, as in the 
standard dimensional regularization\cite{bolini}. Even integrals pertaining to 
the trickier non-covariant gauges, such the light-cone\cite{probing} and
Coulomb\cite{coulomb} ones, can be performed using the same approach
(however, we will deal with them in another work). In the following 
sections we will show how this can be done.

The outline for our paper is as follows: in section II we consider the
simplest two-loop diagram, our workhorse, as a pedagogical example and 
to present, in a clean way, the technique of negative-dimensional 
integration. Section III is devoted to the referred diagram but now with
tensorial structure. The \ndim{} can handle vector, second rank tensor and
higher integrals, all at the same time. In the fourth section we replace
massless propagators  by massive ones and in section V, we consider the
massless diagram with N-insertions of the same type. In the last section, VI,
we present our concluding  remarks.

\section{Simplest two-loop diagram.}

To make things clear we begin with the diagram of figure 1. In a 
massless
scalar theory it is represented by,

\be A =  \int \frac{d^D\! q\; d^D\! k}{(q^2)(k^2)(p-q-k)^2} .\ee

%% Troca de pagina!!!!!!!!!!!!!!!!!
\vfill\eject
{\it Negative-dimensional approach.} 

Consider the gaussian integral,
\be\label{gauss} G =  \int d^D\!\! q \; d^D\! k\;\;\exp{\left[-\alpha 
q^2
-\beta k^2 -\gamma (p-q-k)^2\right]}, \ee
where $(\alpha,\beta,\gamma)$ are such that $G$ is well-behaved. We will 
see
that it is the generating functional of negative-dimensional integrals, 
see
eq.(\ref{taylor}). Integrating over $q$ and $k$ is very easy, 
\be\label{g} G =  \left(\frac{\pi^2}{\lambda}\right)^{D/2}\exp{\left(
-\frac{\alpha\beta \gamma}{\lambda}p^2\right)} ,\ee  where $\lambda =
\alpha\beta+ \alpha\gamma + \beta\gamma$. Expanding (\ref{g}) in Taylor 
series
and using a multinomial expansion for $\lambda$ we get,

\be\label{serie} G =  \pi^D \sum_{n_1,n_2,n_3,n_4= 0}^\infty
\frac{\alpha^{n_{123}} \beta^{n_{124}}
\gamma^{n_{134}}(p^2)^{n_1}}{n_1!n_2!n_3!n_4!}(-n_{1}-D/2)! ,\ee
where due to our multinomial expansion, $n_{234} =  -n_1 - D/2$, and we 
define 
$$ n_{12} =  n_1 + n_2, \qquad\quad n_{123} =  n_1 + n_2 + n_3 , $$ 
and so forth.

On the other hand, Taylor expanding (\ref{gauss}), 
\be\label{taylor} G =  \sum_{i,j,l= 0}^\infty 
\frac{(-1)^{i+j+l}\alpha^i \beta^j
\gamma^l}{i!j!l!} \int d^D\!\! q \; d^D\! k \;\; 
(q^2)^i(k^2)^j(p-q-k)^{2l}
,\ee 
one generates the negative-$D$ integral. Now, comparing (\ref{serie}) 
and
(\ref{taylor}) we solve for the integral above,

\beq\label{geral} {\cal A}(i,j,l) &= & \int d^D\!\! q\;d^D\! k\;\;
(q^2)^i(k^2)^j(p-q-k)^{2l} \\
&= & \label{geral-2}\frac{\pi^D i!j!l!}{(-1)^{i+j+l}}
\sum_{n_1,n_2,n_3,n_4= 0}^\infty 
\frac{(p^2)^{n_1}(-n_{1234}-D/2)!}{n_1!
n_2!n_3!n_4!} \delta_{n_{123},i} \delta_{n_{124},j} \delta_{n_{134},l}
\delta_{n_{234}, -n_1 - D/2},\eeq 
where the Kronecker deltas give rise to a system of 4 equations and 4 
``unknowns''.
Plugging the solution into (\ref{geral-2}) provides us the result of 
${\cal A}(i,j,l)$, namely,

\be\label{facil} {\cal A}(i,j,l) =  \frac{\pi^D \G(1+i)\G(1+j)\G(1+l)
\G(1-\s-D/2)}{\G(1-i-D/2)\G(1-j-D/2)\G(1-l-D/2)\G(1+\s)} (p^2)^\s ,\ee
where $\s= i+j+l+D$. However, the above result is valid on the
negative-dimensional region and positive exponents of propagators
$(i,j,l\geq 0)$. To bring it to our real physical world we must invoke 
the
principle of analytic continuation. This is a quite simple 
step\cite{lab}:
group the gamma functions into Pochhammer symbols and use one of its
properties, \be (a)_k \equiv (a|k) =  \frac{\G(a+k)}{\G(k)}, 
\qquad\qquad
(a|-k) =  \frac{(-1)^k}{(1-a|k)}. \ee
Doing so, we get the result in our positive-dimensional world,

\be\label{escalar} {\cal A}^{AC}(i,j,l) =  \pi^D (p^2)^\s (-i|i+j+D/2)
(-j|j+k+D/2) (-k|i+k+D/2) (\s+D/2|-2\s -D/2),\ee and negative exponents 
of
propagators $(i,j,l)\leq 0$ in Euclidean space. Observe that the result 
is
symmetric in the propagators exponents reflecting the symmetry of figure 
1.

The recipe for calculating Feynman integrals using \ndim{} technology is 
quite
simple: i) to each loop write a Gaussian integral whose arguments
are the propagators of the diagram in question; ii) complete the 
square(s) and
integrate; iii) take the original Gaussian integral, Taylor expand the
exponential and change the order $\sum \leftrightarrow\int$; this 
operation
generates negative-dimensional integrals; iv) the equality of these two
expressions must hold, so one can solve for the negative-dimensional 
integral
and gets $n$-fold series involving Kronecker deltas; v) such Kronecker 
deltas
give rise to a system of linear algebraic equations, in most cases it 
has not
a unique solution, since it is a rectangular matrix\cite{box}; vi) 
Plugging the
solution(s) into that series representation provides the result of
negative-dimensional integral -- sometimes in massless cases there 
appears
degenerate solutions\cite{lab}; vii) analytically continue the referred
result to positive-dimensional region using the above property of 
Pochhammer
symbols. The whole procedure is quite simple, and we will show that in 
cases
of interest. 

\section{Tensorial structure.}

The previous result was obtained with amazing easiness: arbitrary 
negative
exponents of propagators, positive dimension and no numerical 
calculations.
The reader can rightfully ask: Does \ndim{} work for tensorial 
numerators as
well? The answer is yes\cite{tensor}. We need to modify only one thing.

Consider the integral,

\be {\cal B}(i,j,l,m) =  \int d^D\!\! q\; d^D\! k \;\;
(q^2)^i(k^2)^j(p-q-k)^{2l} (2q\cdot p)^m .\ee

From,
\beq\label{gauss-tens} G_T &= & \int d^D\!\! q \; d^D\! k
\;\;\exp{\left[-\alpha q^2 -\beta k^2 -\gamma (p-q-k)^2 -2\phi q\cdot
p\right]} \\ 
&= & \left(\frac{\pi^2}{\lambda}\right)^{D/2}\exp{\left(
-\frac{\alpha\beta \gamma +2\beta\gamma\phi - \beta\phi^2
-\gamma\phi^2}{\lambda} p^2\right)}, \nonumber\\ 
&= & \pi^D \sum_{n_1,...,n_7= 0}^\infty \frac{ (-1)^{n_{12}}
2^{n_2} \alpha^{n_{157}} \beta^{n_{12356}} \gamma^{n_{12467}
\phi^{n_2+2n_{34}}} (p^2)^{n_{1234}}}{n_1!n_2! n_3!n_4! n_5!n_6!n_7!}
(-n_{1234}-D/2)! ,\\  
&= & \sum_{i,j,l,m= 0}^\infty \frac{(-1)^{i+j+l+m}\alpha^i
\beta^j \gamma^l\phi^m}{i!j!l!m!} \int d^D\!\! q\;; d^D\! k
\; (q^2)^i(k^2)^j(p-q-k)^{2l}(2q\cdot p)^m .\eeq

As we did in the previous section, solving for ${\cal B}$, leads to

\beq\label{b-geral}{\cal B}(i,j,l,m) &= & \frac{\pi^D 
i!j!l!m!}{(-1)^{i+j+l+m}}
\sum_{n_1,...,n_7= 0}^\infty \frac{2^{n_2}
(-1)^{n_{12}} (p^2)^{n_{1234}}(-n_{1234}-D/2)!}{n_1! 
n_2!n_3!n_4!n_5!n_6!n_7!}
\delta_{n_{157},i} \delta_{n_{12356},j} \delta_{n_{12467},l} \\
&& \times \delta_{n_{n_2+2n_{34}},m} \delta_{n_{567}, -n_{1234} -
D/2}.\nonumber\eeq 
Observe that now the system generated by the Kronecker deltas does not 
have a
unique solution, since there are seven ``unknowns'' and five equations 
and, by
this very reason, two of them will be left undetermined. There are
$(C^7_5= 7!/5!2!= 21)$ distinct ways of solving this $5\times 7$ 
system, but 5
of them have no solution at all. 

In a previous work\cite{tensor} we did show that all non-trivial 
solutions are
legitimate and lead to the same result for the Feynman integral in 
question.
Here the same occurs -- but we will not prove it. 

Note that the tensorial sector of the Feynman integral ${\cal 
B}(i,j,l,m)$ is
contained in the factor $(2q\cdot p)^m$ and for this very reason the 
exponent
$m$ can not be analytically continued to allow for negative values. In 
other
words, we must invoke the principle of analytic continuation for three
exponents of propagators leaving the fourth, $m$, untouched.

All the solutions will give rise to a double series of hypergeometric 
type,
since we have a 7-fold series and only 5 Kronecker deltas in 
(\ref{b-geral}).
However, from the theory of hypergeometric series\cite{luke} we know
that when one of its numerator parameters is a negative integer, say 
$-m$, the 
series is truncated and has only $m$ terms. For this reason we will 
consider a
solution that is obtained when $\{n_3,n_4\}$ are left undetermined.

\be\label{sol-tensor} {\cal B}(i,j,l,m) =  g_{\cal B} 
\sum_{n_3,n_4= 0}^\infty
\frac{(-m/2|n_{34}) (1/2-m/2|n_{34}) (D/2+j|n_4) (D/2+l|n_3)}{(1+\s' 
-m|n_{34})
(1-i-m-D/2|n_{34}) n_3!n_4!} ,\ee 
where 
\be g_{\cal B} =  \frac{(-\pi)^D (p^2)^{\s'} 2^m \G(1+i)\G(1+j)\G(1+l)
\G(1-\s'-D/2)}{\G(1-j-D/2)\G(1-l-D/2)\G(1-i-m-D/2) \G(1+\s'-m) }   ,\ee
$\s'= \s+m$ and we use the relation\cite{luke},

$$ (a|2b) =  2^{2b} (a/2|b)(1/2+a/2|b) .$$
Note that for positive $m$, which is the relevant condition, the
series (\ref{sol-tensor}) is always truncated (when it is even the
first factor in the numerator is the responsible for that, whereas for 
$m$ odd
the second factor truncates the series.) Analytically continuation of 
$g_{\cal
B}$ gives, \be g_{\cal B}^{AC} =  \pi^D (p^2)^{\s'} 2^m (-i|-j-l-D) 
(-j|j+l+D/2)
(-l|j+l+D/2) (\s'+D/2|i+m-\s') ,\ee
and the final result, in positive dimension, is given by the series in 
equation
(\ref{sol-tensor}) times $g_{\cal B}^{AC}$,

\be\label{result} {\cal B}^{AC}(i,j,l,m) =  g_{\cal B}^{AC}
\sum_{n_3,n_4= 0}^\infty \frac{(-m/2|n_{34}) (1/2-m/2|n_{34}) 
(D/2+j|n_4)
(D/2+l|n_3)}{(1+\s' -m|n_{34}) (1-i-m-D/2|n_{34}) n_3!n_4!} .\ee 

Observe that for $m= 0$ we obtain the scalar case (\ref{escalar}), for 
$m= 1$ an integral with vector numerator, for $m= 2$ second rank tensor and so 
forth. The results, of course, are  contracted with the external momentum 
$p^\mu$. The astonishing point is that all these {\bf new} results are
contained {\it in the  same formula}, namely equation (\ref{result}).

\section{Massive propagators.}

\ndim{} is a powerful technique. It gives, simultaneously, vector, 
second rank
tensor and higher order integrals. A second question one could ask is: 
Does 
\ndim{} work for massive propagators as well? The answer is also yes and 
we
need to do only slight modifications.

Let our generating function, corresponding to diagram of figure 1 where 
now
the virtual particles have distinct masses, be

\beq\label{gauss-m} G_m &= & \int d^D\!\! q \; d^D\! 
r\;\;\exp{\left\{-\alpha
(q^2-m_1^2) -\beta (r^2-m_2^2) -\gamma \left[(p-q-r)^2-
m_3^2\right]\right\}}
\\
&= & \sum_{i,j,k= 0}^\infty \frac{(-1)^{i+j+k}\alpha^i \beta^j
\gamma^k}{i!j!k!} {\cal M}(i,j,k) \nonumber\eeq
where,
\be {\cal M}(i,j,k) =  \int d^D\!\! q \; d^D\! r\;\;
(q^2-m_1^2)^i(r^2-m_2^2)^j\left[(p-q-r)^2-m_3^2\right]^k\\ , \ee
using (\ref{g}) we get,

\be\label{gm} G_m =  \exp{\left( \alpha m_1^2 + \beta m_2^2 +\gamma 
m_3^2 
\right)} G ,\ee
and following the general procedure, as in the previous cases, one can 
write the
integral as,
\be {\cal M}(i,j,k) =  \frac{\pi^Di!j!k!}{(-1)^{i+j+k}}
\sum_{n_1,...,n_7= 0}^\infty \frac{(-n_4-D/2)! (m_1^2)^{n_1}
(m_2^2)^{n_2}(m_3^2)^{n_3} (-p^2)^{n_4} 
}{n_1!...n_7!}\delta_{n_{1456},i}
\delta_{n_{2457},j} \delta_{n_{3467},k} \delta_{n_{4567},-D/2}. \ee  
In this case, the Kronecker deltas give rise to a $4\times 7$ system of 
linear algebraic equations. We have 35 possible solutions for such system but 
15 of them have no solution at all. So, we are left with 20 triple series, 
following the prescription of summing the ones which have the same
variables\cite{box,probing} (it is equivalent to say that we sum the 
ones which have the same region of convergence\cite{lab}), we get four 
possible triple series (of hypergeometric type) representing the Feynman
integral ${\cal M}(i,j,k)$,

$$ \left(\frac{m_1^2}{p^2}\right)^a \left(\frac{m_2^2}{p^2}\right)^b
\left(\frac{m_3^2}{p^2}\right)^c , \left(\frac{m_1^2}{m_3^2}\right)^a
\left(\frac{m_2^2}{m_3^2}\right)^b  \left(\frac{p^2}{m_3^2}\right)^c , 
\left(\frac{m_1^2}{m_2^2}\right)^a \left(\frac{m_3^2}{m_2^2}\right)^b 
\left(\frac{p^2}{m_2^2}\right)^c, 
\left(\frac{m_2^2}{m_1^2}\right)^a \left(\frac{m_3^2}{m_1^2}\right)^b 
\left(\frac{p^2}{m_1^2}\right)^c. $$
For the first one we have eight solutions, the second, third and fourth 
have four, so we have 20 possible solutions of the system generated by the
Kronecker deltas $(8+4+4+4=20)$. Since the last three have the same 
form, due to the symmetry of the diagram in question, we will consider only
one of  them; the others can be obtained changing masses and exponents of
propagators.

We quote only the results, where the analytic continuation process has 
been already carried out. The first triple series is,

\beq {\cal M}_1(i,j,k,\{z\}) &= & \left[ f_1 {\cal
F}_C^{(3)}(-k,1-k-D/2;1+i+D/2,1+j+D/2,1-k-D/2) + (j\leftrightarrow k) 
+(i\leftrightarrow k)\right] \nonumber\\ &&+ \left[ f_2 {\cal 
F}_C^{(3)}(-j-k-D/2,1-j-k-D;1+i+D/2,1-j-D/2,1-k-D/2) + (i\leftrightarrow 
j) +(k\leftrightarrow i)\right] \nonumber\\
&&+ f_3 {\cal F}_C^{(3)}(-\sigma,1-\sigma-D/2;1-i-D/2,1-j-D/2,1-k-D/2),\eeq  
where  

$$ f_1 =  (-\pi)^D (m_1^2)^{i+D/2}(m_2^2)^{j+D/2} (p^2)^k (-i|-D/2) 
(-j|-D/2)
,$$ 

$$ f_2 =  (-\pi)^D (m_1^2)^{i+D/2} (-p^2)^{j+k+D/2} (-i|-D/2) 
(-j|-k-D/2)
(j+k+D|-k-D/2) (-k|2k+D/2) ,$$ 

$$ f_3 =  (-\pi)^D (p^2)^\sigma (-i|2i+D/2) (-j|2j+D/2)(-k|2k+D/2) 
(D/2|-\sigma
-D/2) .$$
Note that one of the solutions has a factor 
$(0|D/2)= \Gamma(D/2)/\Gamma(0)$, so that we have in fact seven terms 
(in positive
dimension) instead of eight (in negative dimension.) The second triple 
series
of hypergeometric type is represented by a sum of four terms,

\beq {\cal M}_2(i,j,k,\{w\}) &= & g_1 {\cal 
F}_C^{(3)}(-k,D/2;1+i+D/2,1+j+D/2,
D/2) \nonumber\\
&&+ g_2 {\cal F}_C^{(3)}(-j-k-D/2, -j; 1-j-D/2, D/2, 1+i+D/2)\nonumber\\
&& + g_3 {\cal F}_C^{(3)}(-i, -i-k-D/2; 1-i-D/2, 1+j+D/2, D/2) \\
&& + g_4 {\cal F}_C^{(3)}(-\sigma,-i-j-D/2; 1-i-D/2, 1-j-D/2, 
D/2)\nonumber
,\eeq  
where the hypergeometric series which appears in the above results 
\be {\cal F}_C^{(3)}(\alpha,\beta;\gamma,\theta,\phi,\{x\}) = 
\sum_{n_1,n_2,n_3= 0}^\infty \frac{(\alpha|n_{123}) (\beta|n_{123}) 
x_1^{n_1}
x_2^{n_2} x_3^{n_3}}{n_1!n_2!n_3!(\gamma|n_1) (\theta|n_2) (\phi|n_3)} 
,\ee
is a Lauricella function\cite{berends} which converges if,
\be \left|x_i\right|<1, \qquad {\rm and} \qquad\quad \sqrt{|x_1|} +
\sqrt{|x_2|} + \sqrt{|x_3|} <1 ,\ee and we define,

$$ g_1 =  \frac{(-\pi)^D}{ (-1)^{i+j+k}}
(m_1^2)^{i+D/2}(m_2^2)^{j+D/2}(m_3^2)^{k}(-i|-D/2)(-j|-D/2),$$

$$ g_2 =  \frac{(-\pi)^D}{ (-1)^{i+j+k}}(m_1^2)^{i+D/2} 
(m_3^2)^{j+k+D/2}
(D/2|j)(-i|-D/2) (-k|-j-D/2) ,$$

$$ g_3 = \frac{(-\pi)^D}{ (-1)^{i+j+k}}
(m_2^2)^{j+D/2}(m_3^2)^{i+k+D/2}(-j|-D/2)(-k|-i-D/2)
(D/2|i),$$

$$ g_4 =  \pi^D(-m_3^2)^{\sigma}(-i|i+j+D/2)(-j|i+j+D/2) (D/2|-i-j-D)
(-k|-i-j-D) .$$

The variables in ${\cal M}_1$ and ${\cal M}_2$ are respectively $\{x\} 
=  \{z\}$ and $\{x\} =  \{w\}$,
where
$$ z_1= m_1^2/p^2,\qquad\quad z_2= m_2^2/p^2,\qquad\quad 
z_3= m_3^2/p^2, $$

$$ w_1= m_1^2/m_3^2,\qquad\quad w_2= m_2^2/m_3^2,\qquad\quad 
w_3= p^2/m_3^2. $$

Observe that our results, which were obtained simultaneously, agree with
Berends'\cite{berends} {\it et al} ones known in the literature (which 
was obtained by analytic continuation.) In their work, ${\cal 
M}_2(i,j,k,\{w\})$ was calculated first and then analytically continued to
allow other  values of momenta and masses, resulting in ${\cal
M}_1(i,j,k,\{z\})$. However, if the \ac{} formula was not known the result in
the other region would be difficult to obtain. On the other hand, \ndim{}
provides {\it all} the results simultaneously, allowing us to obtain also 
new analytic continuation formulas\cite{probing,box}.

 \section{Insertions.}

A third question the reader could pose is: And for higher number
of loops does \ndim{} accomplish also good results? Our main objective 
is to apply \ndim{} to higher loops, but up to now we are able to ``glue'' 
diagrams with two external legs only.

Consider the diagram of figure 2. It has 4-loops and is represented by,

\be {\cal C}(i,j,k,m,n,s) =  \int d^D\! q\; d^D\! r\; d^D l\;
d^D\! t\;\; (q^2)^i (r^2)^j (p-q-r)^{2k} (l^2)^m (t^2)^n (r-l-t)^{2s} 
,\ee
in a massless theory. Observe that the integrals in $l$ and $t$ are 
equal, see eq.(\ref{geral}), to ${\cal A}(m,n,s;p^2\rightarrow r^2)$.
Rewriting we  get,
\beq {\cal C}(i,j,k,m,n,s) &= & \int d^D\! q\; d^D r\; (q^2)^i (r^2)^j
(p-q-r)^{2k} {\cal A}(m,n,s;p^2\rightarrow r^2), \\
&= &  \pi^D (-m|m+n+D/2)(-n|n+s+D/2)(-s|m+s+D/2)(\s_1+D/2|-2\s_1 -D/2) 
{\cal
A}(i,j+\s_1,k), \nonumber\\
&= & \pi^{2D} (p^2)^{\s_1+\s_2} (-i|i+j+\s_1+D/2) 
(-j-\s_1|j+\s_1+k+D/2)
(-k|i+k+D/2)\nonumber\\
&&\times  (\s_2+D/2|-2\s_2-D/2) (-m|m+n+D/2) (-n|n+s+D/2) (-s|m+s+D/2)
\nonumber\\
&&\times (\s_1+D/2|-2\s_1 -D/2), \eeq 
where $\s_1= m+n+s+D$, $\s_2= i+j+k+D$. 

Let us study the diagram of figure 3, namely the one that has a
single propagator replaced $N$-times by the graph of figure 1.

\be {\cal C}_N(\nu_1,\nu_2,...,\nu_{3N}) =  \int 
\dots\int\prod_{i= 1}^{i= N}
d^D\! q_i\; d^D\! r_i\; (q_i^2)^{\nu_{3i-2}} (r_i^2)^{\nu_{3i-1}}
\left[(r_{i-1}-q_i-r_i)^2\right]^{\nu_{3i}} , \ee
where $r_0^\mu= p^\mu$ is the external momentum and $N$ is the number 
of
insertions. When $N= 1$ the above integral reduces to ${\cal
A}(\nu_1,\nu_2,\nu_3)$.

Applying the above procedure we can easily solve such scalar integral,

\be {\cal C}^{AC}_N(\nu_1,\nu_2,...,\nu_{3N}) =  {\cal
A}^{AC}(\nu_1,\nu_2,\nu_3) {\cal A}^{AC}(\nu_4,\nu_5 + \s_1,\nu_6) 
\times
...\times {\cal A}^{AC}(\nu_{3N-2},\nu_{3N-1}+\s_1+...+\s_{N-1},\nu_{3N})
,\ee where $\s_N =  \nu_{3N-2} + \nu_{3N-1} + \nu_{3N} +D$.

{\it Water melon diagram. Massless case.} 

Recently, the water melon diagram was considered in massive case, 
whether in 
four dimensions and analytic result (see first reference in 
\cite{berends}) or
integral representations of it suitable for numerical calculations (see 
third
reference in \cite{berends}). 

Applying \ndim{} we can solve, exactly, such water melon diagrams, see 
figure
4. Here, we will consider the scalar massless case. At two-loop level 
the
referred water melon is the graph of figure 1, our setting Sun diagram. 
At
three-loop level we begin with,
\beq G_{WM} &= & \int d^D\! q\; d^D\! r\; d^D k\;\; \exp{\left[-\alpha 
q^2
-\beta r^2 -\gamma k^2 -\theta (p-q-r-k)^2\right]}, \\
&= & \left(\frac{\pi^3}{\zeta}\right)^{D/2}
\exp{\left(-\frac{\alpha\beta\gamma\theta}{\zeta}p^2 \right)} ,\eeq
where $\zeta =  \alpha\beta\gamma + \alpha\beta\theta + 
\alpha\gamma\theta +
\beta\gamma\theta$. After a little bit of algebra we get a $4\times 4$ 
system
which gives 

\beq {\cal W}_3(i,j,l,m) &= & \int d^D\! q\; d^D\! r\; d^D k\; (q^2)^i 
(r^2)^j
(k^2)^l (p-q-r-k)^{2m}, \\ 
&= & \frac{\pi^{ 3D/2} i!j!l!m! \Gamma(1-\rho_3-D/2) (p^2)^{\rho_3} }{
(-1)^{i+j+l+m} \Gamma(1-i-D/2) \Gamma(1-j-D/2)\Gamma(1-l-D/2) 
\Gamma(1-m-D/2)\Gamma(1+\rho_3)}  ,\eeq 
in negative dimension and Euclidean space. We define 
$\rho_3= i+j+l+m+ 3D/2$.
Generalizing this result to $N$-loops is quite easy,

\beq\label{melao} {\cal W}_N(\{\nu_n\}) &= & \int\dots\int 
\prod_{i= 1}^{i= N}
d^D\! q_i\; (q_i^2)^{\nu_i} (p-q_1-q_2-\dots -q_N)^{2\nu_{N+1} }, \\  
&= & \frac{\pi^{ND/2} \nu_1!\dots \nu_{N+1}! \Gamma(1-\rho_N-D/2)
(p^2)^{\rho_N} }{ (-1)^{\Sigma \nu} \Gamma(1-\nu_1-D/2)\dots
\Gamma(1-\nu_{N+1}-D/2) \Gamma(1+\rho_N)}  ,\eeq  
where $\Sigma \nu =  \nu_1 +\dots + \nu_{N+1} $ and $\rho_N= 
\Sigma\nu + ND/2$. Carrying the analytic continuation out we get,

\be \label{melao-ac} {\cal W}^{AC}_N(\{\nu_n\}) = 
\pi^{ND/2}(p^2)^{\rho_N}
(-\nu_1|2\nu_1+D/2) (-\nu_2|2\nu_2+D/2) \dots (-\nu_N|2\nu_N+D/2) 
(\rho_N
+D/2| -2\rho_N-D/2) ,\ee 
the result for negative exponents of propagators and positive dimension. As
far as we know, this result was not known for arbitrary exponents of
propagators. Observe that when one is allowed to  ``gluing'' diagrams, the
result can easily be generalized to N-loops.

\section{Conclusion.}\ndim{} is a suitable technique to tackle the task 
of calculating multiloop Feynman  integrals. Massless, massive, scalar,
tensorial, even the ones for non-covariant gauges, such as the light-cone 
gauge\cite{probing,coulomb} are easily performed. In all of them exponents of
propagators are left arbitrary as well as the dimension $D$, just like in
plain dimensional regularization. Usual parametric integrals are replaced by
one Gaussian integral over each momentum flowing in the loop, and the main
task is to solve systems of linear algebraic equations. Another point that we
would  like to stress is that no numerical calculations are required at all.

\acknowledgments{A.G.M.S. gratefully acknowledges FAPESP (Funda\c c\~ao 
de Amparo \`a Pesquisa do Estado de S\~ao Paulo, Brasil) for financial
support.}  \end{document}